\documentclass[letterpaper]{rmf-d}
\usepackage{nopageno,rmfbib,multicol,times,amsmath,amssymb}
\usepackage[latin1]{inputenc}
\usepackage[spanish,english]{babel}
\usepackage[dvips]{graphicx}
\usepackage{subfigure}
\usepackage[hang]{caption2}
\usepackage{natbib}
\def\rmfcornisa{ \hfill\rmf\ \hfill }
\def\rmfcintilla{{\it Rev.\ Mex.\ F\'{\i}s.\/}} 
\newcommand{\ia}{\'{\i}}

\newcommand{\zsun}{$Z_{\odot}$}

\renewcommand{\thesubfigure}{\arabic{subfigure}}
\makeatletter
\renewcommand{\@thesubfigure}{Fig. \thesubfigure:\space}
\renewcommand{\p@subfigure}{}

\clearpage
\rmfcaptionstyle
\begin{document}
\title{Spectroscopic Dating of Stellar Populations from Local Star Clusters to
Distant Galaxies and the Age of the Universe
\footnote
{
To appear in {\it Proceedings of the 3rd Congress of the Venezuelan\break
Physical Society}, Revista Mexicana de F\ia sica, 2002}
}
\author{Gustavo Bruzual A.}
\address{Centro de Investigaciones de Astronomía, Apartado Postal 264, Mérida
5101-A, Venezuela}

\maketitle

\recibido{10 de diciembre de 2001}{\today}

\begin{abstract}
I summarize in terms of evolutionary population synthesis models our current
understanding of stellar populations in different environments, from star
clusters in nearby galaxies to distant galaxies in clusters or seen through
gravitational lenses.
Constraints on the age of the stellar populations in clusters and galaxies
derived from comparing predicted and observed spectral energy distributions are
examined and contrasted with the maximum ages allowed by cosmological models
at the corresponding redshift.
\end{abstract}
\vspace{-2pt}
\keys{galaxies: general, galaxies: evolution, galaxies: stellar populations}

\begin{resumen}
Presento un resumen del conocimiento actual de las propiedades de las
poblaciones estelares presentes en diferentes ambientes, desde cúmulos
estelares en galaxias cercanas hasta galaxias lejanas en cúmulos de galaxias o
galaxias observadas gracias a lentes gravitacionales, en términos de modelos
de s\ia ntesis evolutiva de poblaciones estelares.
Se examinan las restricciones que se pueden imponer en la edad de las
poblaciones estelares en c\'umulos estelares y galaxias a partir de la
comparaci\'on de espectros predichos y observados, y se contrastan con la
m\'axima edad permitida por modelos cosmológicos al correspondiente valor del
corrimiento al rojo.
\end{resumen}
\vspace{-2pt}
\descript{galaxias: general, galaxias: evolución, galaxias: poblaciones estelares}



\section{Introduction}
Evolutionary population synthesis provides an important tool to study the
stellar content of star clusters and galaxies as a function of age.
The theory of stellar evolution makes definitive predictions concerning
the bolometric luminosity and effective temperature of a star of a given
mass and chemical composition.
As time proceeds, the function describing the radial dependence of the chemical
composition of the material inside the star varies, resulting in changes in the
bolometric luminosity and the effective temperature characterizing a star of a
given initial mass.
The evolutionary track for this stellar mass is the line joining these
successive points in the bolometric luminosity-effective temperature plane,
known as the Hertzsprung-Russell diagram (HRD).
Complete sets of evolutionary tracks for stars of a wide mass range and various
initial metal contents covering all significant stellar phases are available
in the literature (Fagotto et al. 1994a, b, c; Girardi et al. 2000). 
Assuming an initial mass function (IMF, e.g. Salpeter 1955), we can compute
the number of stars of a given mass born at time zero, and then follow the
evolution of this population in the HRD using a specific set of evolutionary
tracks.
Knowledge of the spectral energy distribution (SED) at each position in the
HRD visited by the stars during their evolution, allows us to compute the
integrated SED for this initial-burst or simple stellar population (SSP) as a
function of time.
By means of a convolution integral (Bruzual \& Charlot 1993) the evolution
of the SED can be computed for an arbitrary star formation rate (SFR) and 
a chemical enrichment law.
A detailed description of the evolutionary synthesis technique has been
published by Bruzual (1999). A discussion of the most important results
obtained from spectral evolutionary models is also presented in this paper,
as well as in Bruzual (2001a,b).

Several stellar spectral libraries are currently available. The Pickles (1998)
atlas provides good coverage of the HRD for stars of solar metallicity (\zsun)
at medium spectral resolution. Recently, Le Borgne et al. (2002a) have compiled
an equivalent atlas at 3\AA\ spectral resolution (1\AA\ sampling) which is
complete for stars of solar metallicity, but includes a large number of spectra
of non-solar metallicity stars. On the theoretical side the compilation by
Lejeune et al. (1997, 1998) and Westera et al. (2002) provide libraries of model
atmospheres for stars of various metallicities but at $\approx$20\AA\ spectral
resolution. The last two libraries are largely based on the Kurucz (1995)
series of model stellar atmospheres.

In a parallel paper (Bruzual 2002), I have discussed the advantages of using
a high resolution stellar spectral library, such as the Le Borgne et al. (2002a)
atlas, in a set of evolutionary population synthesis models (Bruzual \& Charlot
2002, BC02 hereafter). These models preserve the properties of models built at
lower spectral resolution in what respects to integrated photometric properties
of the stellar population, e.g. luminosity and color evolution, but achieve a
much greater level of detail in reproducing the spectral features of the
integrated population. This is particularly important when comparing model
SEDs with observed spectra obtained with the high resolution spectrograph
available in the Hubble Space Telescope (HST) and in the new generation of
ground based high performance optical telescopes (Keck, VLT, Sloan).
In this paper I show the results of comparing the predictions of the BC02 models
with the observed SED of star clusters and galaxies seen at various cosmological
epochs, as measured by their redshift ($z$).

\begin{figure}
\begin{center}
\subfigure[Comparison of the observed SED of the young massive cluster W3 in
the galaxy NGC 7252 (kindly provided by F. Schweizer, heavy line) with a
$Z =$ \zsun\ SSP at 0.57 Gyr computed
for the Salpeter (1955) IMF using the Le Borgne et al. (2002a) stellar atlas
(thin line). The model SED is the line extending to the extreme left and right
in the plot.]
{\includegraphics[scale=.65,angle=270]{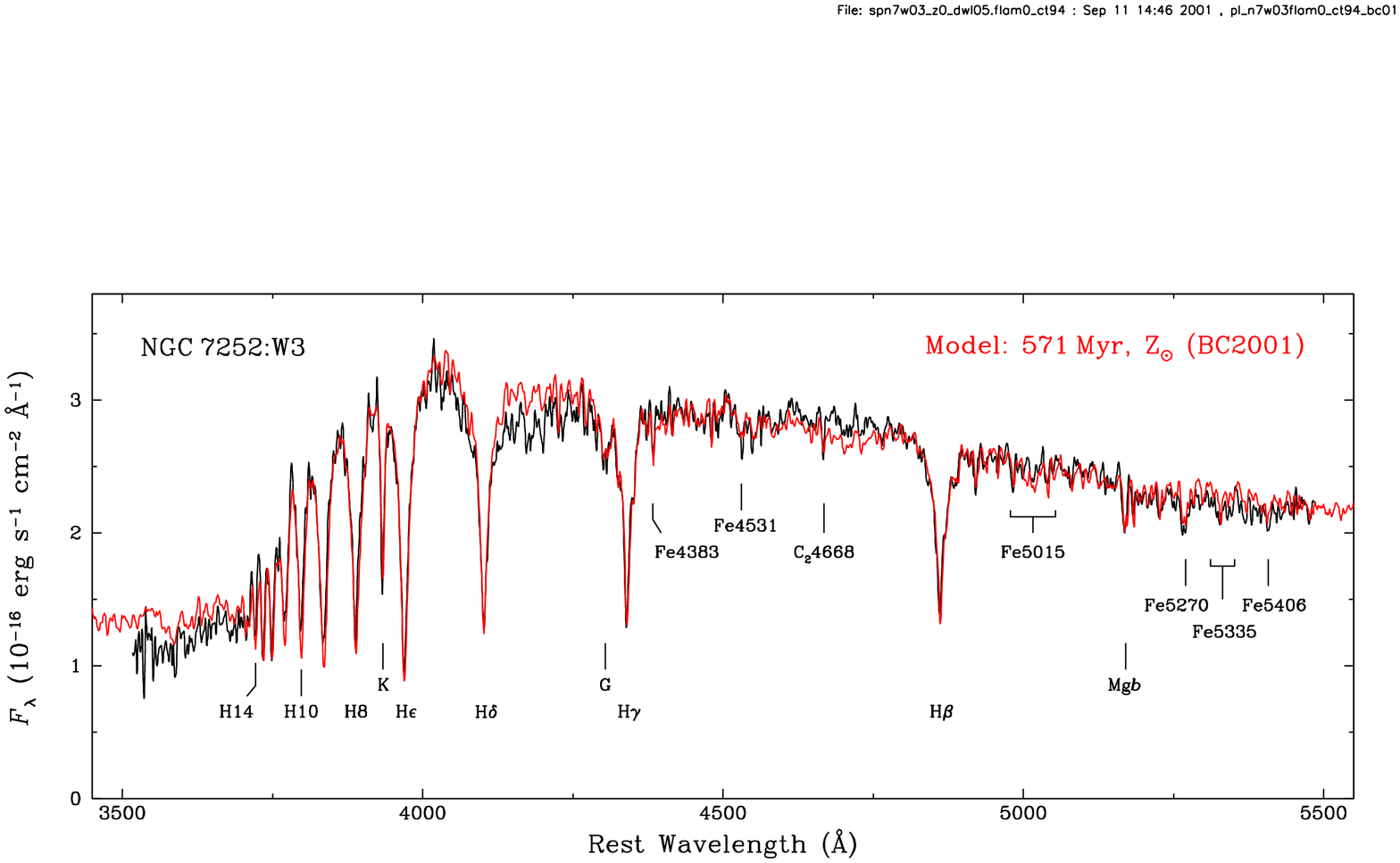} }
\end{center}
\end{figure}

\section{Spectroscopic Age of Stellar Populations}

In this section I use the BC02 models to derive the spectroscopic age that
characterizes the observed SED of several sources at different distances from
us. A model consists of a spectrum which evolves in time from age zero to 20
Gyr in 221 time steps. By minimizing
$\chi^2 = \sum [ log~ F_\lambda(observed) - log~ F_\lambda(model)]^2$,
we derive for each problem SED the (spectroscopic) age at which the model
matches more closely the observed spectrum.

\subsection{Star Clusters}

Fig 1 shows a comparison of a BC02 model SED with the observed spectrum of
the young massive (globular) cluster W3 in the galaxy NGC 7252, kindly
provided by
F. Schweizer. The model corresponds to a $Z =$ \zsun\ SSP at 0.57 Gyr computed
for the Salpeter (1955) IMF using the Le Borgne et al. (2002a) stellar atlas.
The spectra show a remarkable degree of agreement in the shape and intensity
of the Hydrogen Balmer lines (H$_\beta$ through H14), the Mgb feature, the Fe
and other lines indicated in the plot. The intensity of the Balmer lines is an
excellent indicator of the age of the stellar population, provided that the
spectral resolution of both spectra match in this wavelength range. This is
the case in Fig 1. Models built with lower resolution libraries do not constrain
the age of the cluster population so accurately.

\subsection{Nearby Galaxies}

In Fig 2 I show a VLT spectrum of an elliptical galaxy at $z = 0.142$
(kindly provided by J.-F. Le Borgne) in the
field of the cluster of galaxies AC114 together with a BC02 model that produces
the best match to the observed SED at an age of 12 Gyr. The model is the same
one shown in Fig 1, but seen 11.43 Gyr later in its passive evolution.
Many of the spectral features seen in the problem galaxy SED are real since
they are also present in the stellar spectra used to build the model.

\subsection{Intermediate Distance Galaxies}

Stockton (2001) has identified galaxies dominated by old stellar populations
at moderately high redshifts ($z \approx 1.5)$.
Presumably, these galaxies formed most of their stars early in the history of
the universe and have evolved passively later on.
The data points in the different frames of Fig 3 show the broad band
photometry in the galaxy rest frame for LBDS 53W091 (Spinrad et al. 1997), and
six galaxies from Stockton's sample.
The colors or broad band fluxes of LBDS 53W091 are well reproduced by the same
solar metallicity, Salpeter IMF, BC02 model SSP introduced above at an age of
1.4 Gyr (top frame on the left hand side in Fig 3; see also Bruzual \& Magris
1997).
Two of the galaxies observed by Stockton have the colors of this population at
2 Gyr, whereas the remaining 4 galaxies seem older, $\approx 4 Gyr$.
The 1.4 Gyr SED is repeated in all the frames as a guide to the eye in
establishing the differences between the SEDs.
Despite the photometric errors and the intrinsic uncertainties in the population
synthesis technique, it seems safe to conclude that even though these 7 galaxies
are at very similar redshifts, their stellar populations differ in age by a
factor of two.
This may give us an indication on how precisely tuned
star formation was at the time of formation of these galaxies.

\subsection{Distant Galaxies. A Passive Evolution Sequence}

Fig 4 shows in a concise way a comparison of the predicted passive evolution
of the SED of a SSP at 6 different epochs and observations
(either broad band fluxes and/or SEDs) of various galaxies covering a wide
range in redshift space. The age of the model that reproduces the observations
is indicated in each panel.
At one age extreme, object S2 in the top panel of Fig 4, a young (6 Myr)
reddened starburst at $z =1.87$ seen through a gravitational lens (Le Borgne
et al. 2002b), provides an example of a very young population at a cosmological
epoch at which (elliptical?) galaxies dominated by old populations were
already in place. Object H5 (Pell\'o et al. 1999), shown in the second panel
of Fig 4, at $z = 4.05$ is a quite distant galaxy also seen through a
gravitational lens and is characterized by a stellar population with age close
to 60 Myr. Source 3 by Cowie et al. (2001) at $z = 2.6$ is another lensed
galaxy seen at an age close to an order of magnitude older than object H5.
M32 and an average nearby elliptical galaxy SED, shown in the two bottom panels
in Fig 4, characterize present day local counterparts of the spectral energy
distribution of old stellar populations. LBDS 53W091 (Fig 3), reproduced in the
4th panel of Fig 4, provides a link between the younger and older than 1 Gyr
stellar populations.
Fig 4 thus shows a plausible evolutionary sequence of elliptical galaxy spectra.

\begin{figure}
\subfigure[VLT spectrum of an elliptical galaxy at $z = 0.142$ gravitationally 
lensed in the field of
the cluster of galaxies AC114 (kindly provided by J.-F. Le Borgne, heavy line)
compared with the same model of Fig 1 but seen at 12 Gyr (thin line).]
{\includegraphics[scale=.42]{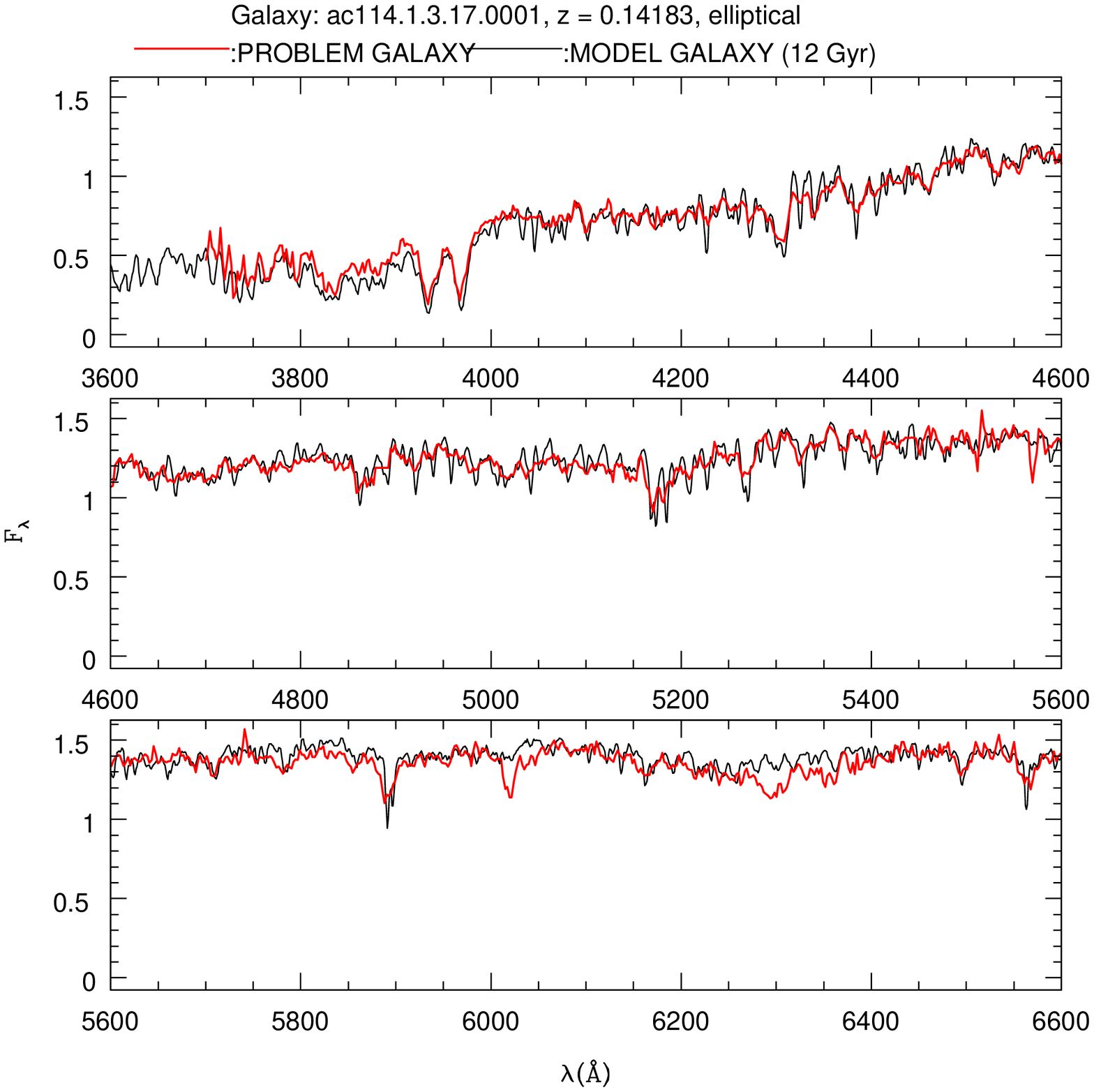} }
\subfigure[Broad band photometry in the galaxy rest frame for LBDS 53W091 and
six galaxies from Stockton's sample compared to the model of Fig 1 at the age
indicated in each panel (thick line). The 1.4 Gyr SED is repeated in all the
frames (thin line) to guide the eye.]
{\includegraphics[scale=.42]{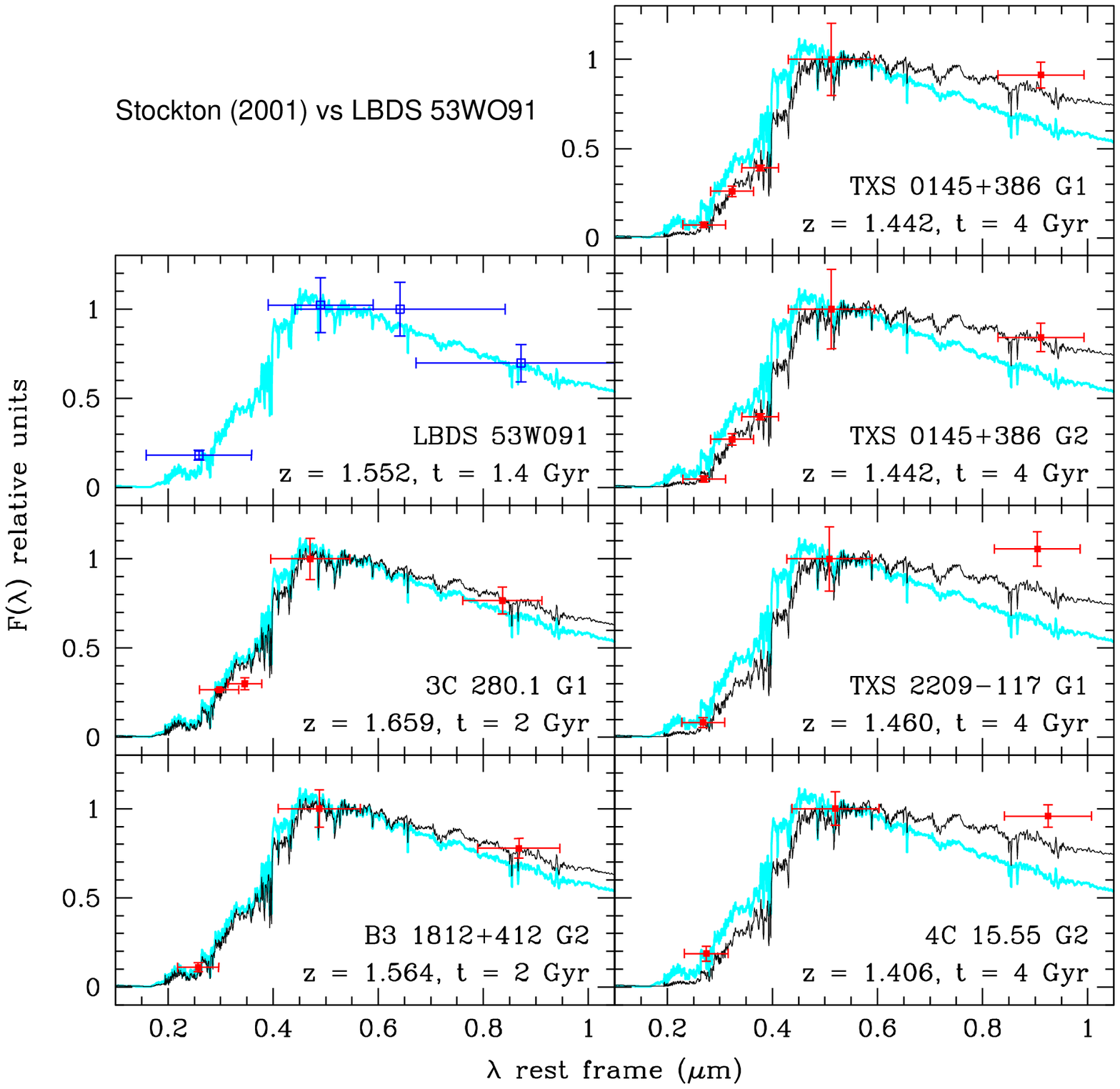} }
\end{figure}

\begin{figure}
\subfigure[Plausible evolutionary sequence of elliptical galaxy spectra.
Comparison of the predicted passive evolution of the SED of a SSP
at 6 different epochs and observations (either broad band fluxes and/or SEDs)
of various galaxies covering a wide range in redshift space. The age of the
model that reproduces the observations is indicated in each panel.]
{\includegraphics[scale=.42]{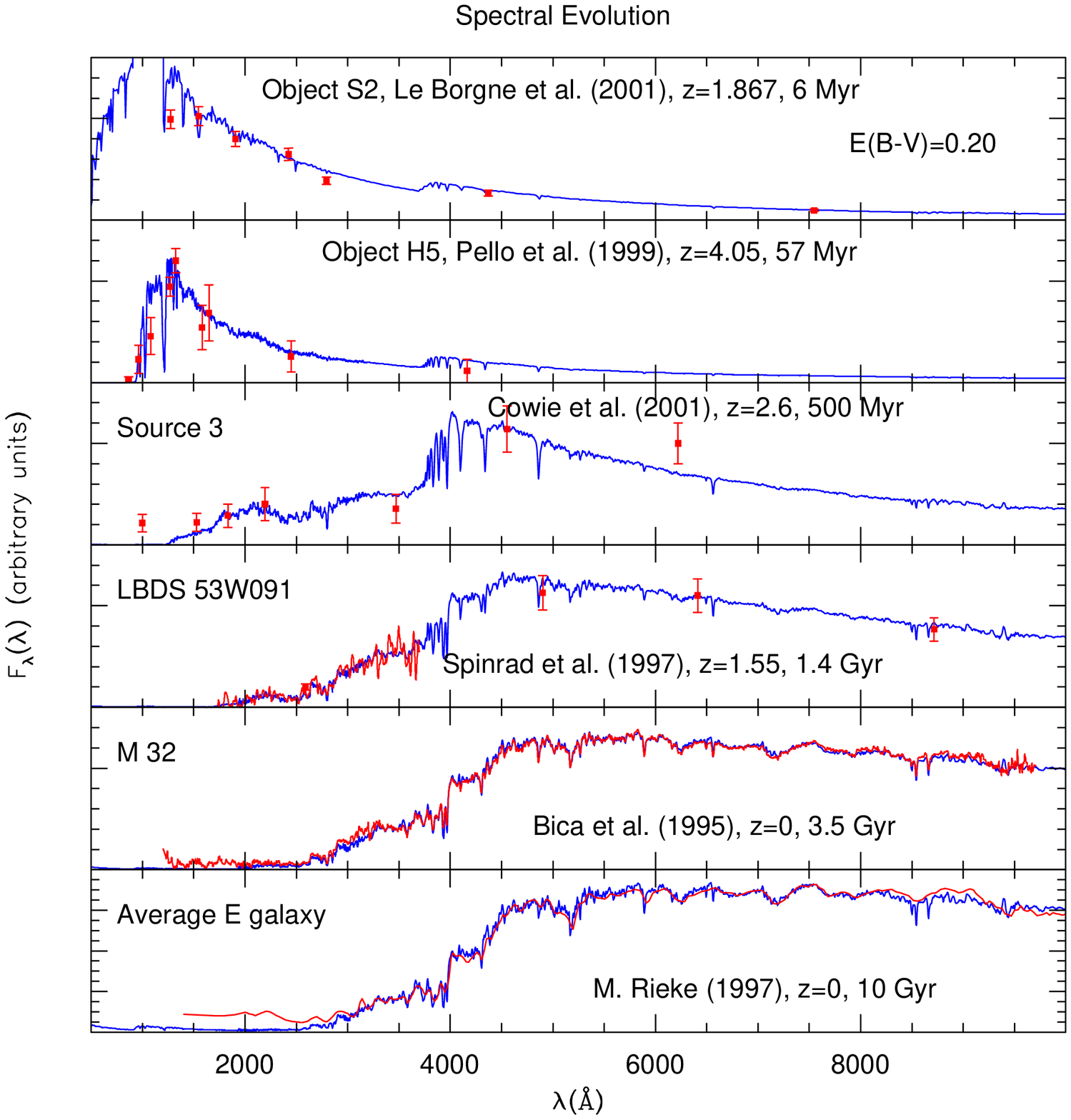} }
\subfigure[Age of the universe as a function of redshift for two different
cosmological models:
$H_0=70$ km s$^{-1}$ Mpc$^{-1}, \Omega_M=1,~\Omega_\Lambda=0,~t_U(0)=9.33$
Gyr (bottom solid line), and
$H_0=70$ km s$^{-1}$ Mpc$^{-1}, \Omega_M=0.28,~\Omega_\Lambda=0.72,~t_U(0)=13.75$
Gyr (top solid line). The data points represent the ages determined
spectroscopically for the galaxies in Fig 4 and globular star clusters taken
from the literature. The open dots represent the galaxies in Stockton's sample
shown in Fig 3. The vertical segments at $z=0$ and $z=1.55$ represent
a range in possible ages and not an error bar.]
{\includegraphics[scale=.42]{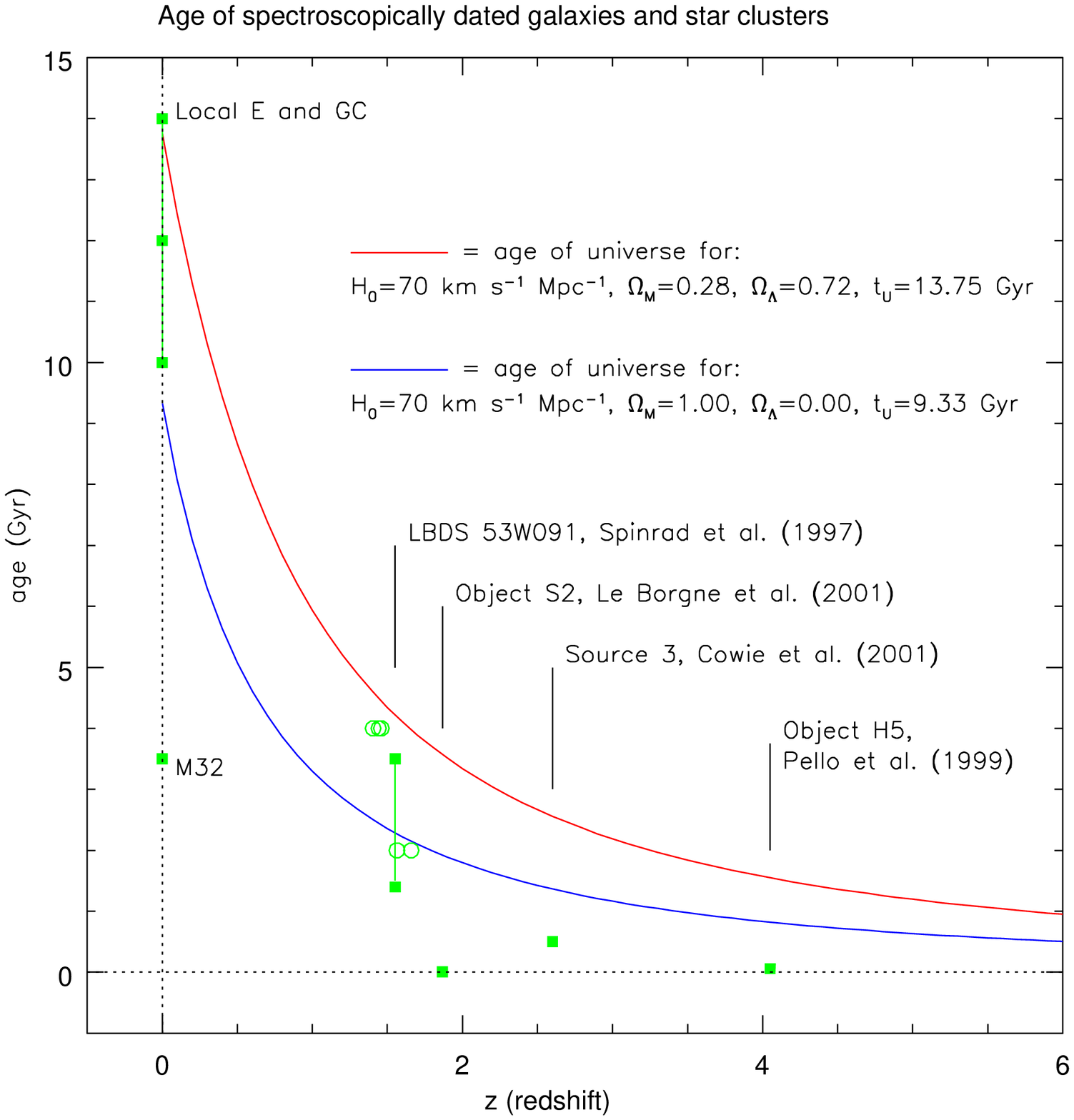} }
\end{figure}

\section{Spectroscopically Dated Galaxies and the Age of the Universe}

Fig 5 shows the age of the universe as a function of redshift for two different
cosmological models.
The data points represent the ages determined spectroscopically for the
different galaxies shown in Figs 3 and 4, as well as globular star clusters taken
from the literature.
It is clear that the age of local E galaxies and globular clusters, and
possibly LBDS 53W091 and two of Stockton's galaxies, rule out the
$H_0 = 70$ km s$^{-1}$ Mpc$^{-1}$, $\Omega_M = 1$, $\Omega_\Lambda = 0$
universe with a maximum age $t_U(0) = 9.33$ Gyr at $z = 0$ (bottom solid line).
On the other hand, the age of these systems and of all galaxies shown in Fig 5,
is below the $age(z)$ line for the standard model,
$H_0 = 70$ km s$^{-1}$ Mpc$^{-1}$, $\Omega_M = 0.28$, $\Omega_\Lambda = 0.72$,
$t_U(0) = 13.75$ Gyr (top solid line).
A selection effect is clear in Fig 5, the sources that have been observed,
except for the galaxies at $z \approx 1.5$,
represent the youngest, and possibly the brightest galaxies at a given $z$,
and may not be representative of the galaxy population as a whole in that
shell of $z$-space.
Equally young stellar populations seem to be present at all $z$'s, at least
from $z\geq 2$ to $z\leq 4$.

\section{Discussion}

I have shown how evolutionary population synthesis models can be used to
assign a spectroscopic age to stellar populations whose SED or multi-broad-band
photometry is available.
The error in the age determination depends on the quality of the observational
data and on the uncertainties in the population synthesis models.
In general, the relative error in the spectroscopic age increases
with decreasing galaxy age.
Despite this caveat, it is possible to establish a plausible sequence along
which the spectra of galaxies may evolve when passive evolution dominates
the evolution of simple stellar populations, at a rate that is consistent
with that expected in the most accepted cosmological model describing our
universe,
$H_0 = 70$ km s$^{-1}$ Mpc$^{-1}$, $\Omega_M = 0.28$, $\Omega_\Lambda = 0.72$,
$t_U(0)=13.75$ Gyr.



\begin{thebibliography}{99}

\bibitem{a}
Bica, E., Alloin, D., Bonatto, C., Pastoriza, M.G., Jablonka, P., Schmidt, A.,
\& Schmitt, H.R. 1996, in {\it A Data Base for Galaxy Evolution Modeling},
eds. C. Leitherer et al., PASP, 108, 996
\bibitem{b}
Bruzual A., G. 1999, in {\it Proceedings of the XI Canary Islands Winter School of Astrophysics on Galaxies at High Redshift}, eds. I. P\'erez-Fournon and F. S\'anchez, Cambridge Contemporary Astrophysics, (in press)
\bibitem{c}
---------. 2001a, Astrophys. \& Space Science 277 (Suppl.), 221
\bibitem{d}
---------. 2001b, in {\it Proceedings of the IAU Symposium No. 207
``Extragalactic Star Clusters''}, eds. D. Geisler and E. Grebel,
Astr. Soc. Pac. Conference Series, (in press)
\bibitem{e}
---------. 2002, Rev. Mex. Astron. Astrophys. Conf. Series, (in press)
\bibitem{f}
Bruzual A., G. \& Charlot, S. 1993, ApJ, 405, 538
\bibitem{g}
---------. 2002, ApJ, in preparation (BC02)
\bibitem{h}
Bruzual A., G. \& Magris, G. 1997, in {\it Proceedings of the STScI May
Symposium ``The Hubble Deep Field''}, eds. M. Livio, S.M. Fall \& P. Madau, p. 9
\bibitem{i}
Cowie, L. L., Barger, A. J., Bautz, M. W., Capak, P., Crawford, C. S., Fabian, A. C., Hu, E. M., Iwamuro, F., Kneib, J.-P., Maihara, T., \& Motohara, K. 2001,
ApJ, 551, L9
\bibitem{j}
Fagotto, F., Bressan, A., Bertelli, G., \& Chiosi, C. 1994a, A\&AS, 100, 647
\bibitem{k}
---------. 1994b, A\&AS, 104, 365
\bibitem{l}
---------. 1994c, A\&AS, 105, 29
\bibitem{m}
Girardi, L., Bressan, A., Bertelli, G., \& Chiosi, C. 2000, A\&AS, 141, 371
\bibitem{n}
Kurucz, R. 1995, private communication
\bibitem{o}
Le Borgne, J.-F., et al. 2002a, A\&A, (in preparation)
\bibitem{p}
Le Borgne, J.-F., Schaerer, D., Bruzual A., G., Pell\'o, R., Lemoine-Busserole,
M., Campusano, L.E., Kneib, J.-P., \& Smail, I., 2002b, A\&A, (in preparation)
\bibitem{q}
Lejeune, T., Cuisinier, F., \& Buser, R. 1997a, A\&AS, 125, 229
\bibitem{r}
---------. 1998, A\&AS, 130, 65
\bibitem{s}
Pell\'o, R., Kneib, J.P., Le Borgne, J.F., B\'ezecourt, J., Ebbels, T.M.,
Tijera, I., Bruzual A., G., Miralles, J.M., Smail, I., Soucail, G., \&
Bridges, T.J. 1999, A\&A, 346, 359
\bibitem{t}
Pickles, A.J. 1998, PASP, 110, 863
\bibitem{u}
Rieke, M. 1997, private communication
\bibitem{v}
Salpeter, E.E. 1955, ApJ, 121, 161
\bibitem{w}
Spinrad, H., Day, H., Stern, D., Dunlop, J., Peacock, J., Jim\'enez, R., \&
Windhorst, R., 1997, ApJ, 484, 581
\bibitem{x}
Stockton, J. 2001, in {\it Astrophysical Ages and Time Scales},
eds. T. von Hippel, N. Manset, and C. Simpson,
Astr. Soc. Pac. Conference Series, Vol. 245, p. 517
\bibitem{y}
Westera, P., Lejeune, T., Buser, R., Cuisinier, F., \& Bruzual A., G. 2002,
A\&A, in press
\end{thebibliography}
\end{document}